\documentclass[preprint,aps,draft]{revtex4}
\usepackage{amsmath,amssymb,amsfonts}

\newtheorem{theorem}{Theorem}
\newtheorem{prop}[theorem]{Proposition}

\newtheorem{exe}[theorem]{Exercise}
\newtheorem{exa}[theorem]{Example}
\newtheorem{defi}[theorem]{Definition}
\newtheorem{remark}{Remark}
\newenvironment{rem}{\begin{remark} \rm}{\end{remark}}

\newenvironment{proof}[1][Proof]{\noindent\textbf{#1.} }{\null\hfill
\rule{0.5em}{0.5em}}
\newcommand{\di}{\mathrm{d}}
\newcommand{\la}{\lambda}
\newcommand{\tr}{\mathrm{tr}\,}

\newcommand{\CS}{{\cal S}}

\newcommand{\CN}{{\cal N}}

\newcommand{\CM}{{\cal M}}
\newcommand{\CQ}{{\cal Q}}

\newcommand{\RR}{{\mathbb R}}

\def\ga{\gamma}         
\def\be{\beta}
\def\al{\alpha}

\def\la{\lambda}        \def\La{\Lambda}

\newcommand{\rref}[1]{(\ref{#1})} 

\def\bih{bi-Ham\-il\-tonian}
\def\varb{\bih\ manifold}

\def\syml{symplectic leaf}

\begin{document}

\title{ Bi-Hamiltonian Aspects 
of a Matrix Harry Dym Hierarchy}

\author{Laura Fontanelli, Paolo Lorenzoni}
\email{laura.fontanelli@unimib.it, paolo.lorenzoni@unimib.it}
\affiliation{
Dipartimento di Matematica e Applicazioni\\
Universit\`a di Milano-Bicocca\\
Via Roberto Cozzi 53, I-20125 Milano, Italy
}%

\author{Marco Pedroni}
\email{marco.pedroni@unibg.it}
\affiliation{
Dipartimento di Ingegneria dell'Informazione e Metodi Matematici\\
Universit\`a di Bergamo\\
Viale Marconi 5, I-24044 Dalmine (BG), Italy
}%

\author{Jorge P. Zubelli}
\email{zubelli@impa.br}
\affiliation{
IMPA\\
Est. D. Castorina 110, Rio de Janeiro RJ 22460-320, Brazil
}%

\date{December 17th, 2007}


\begin{abstract}
We study the Harry Dym hierarchy of nonlinear
evolution equations from the  bi-Hamiltonian view point. This is done by
using the concept of an $\CS$-hierarchy. It allows us to
define a matrix Harry Dym hierarchy of commuting Hamiltonian flows in two fields that
projects onto the scalar Harry Dym hierarchy.
We also show that
the conserved densities of the matrix Harry Dym equation can be found
by means of a Riccati-type equation.
\end{abstract}

\maketitle

\section{Introduction}
An intriguing equation known as the Harry Dym (HD) equation
has attracted the attention of a number of researchers in
integrable systems 
\cite{CD,DK94,GR89,KL94,KL95,OeRo93,RR94,R87,RSV94}.
In one of its incarnations it can be written as
\begin{equation}  \label{bhd1}
 q\sb t=2(1/\sqrt{(1+q)})\sb {xxx}
\end{equation} or equivalently
\begin{equation}  \label{bhd2}
 \rho\sb t=\rho\sp 3\rho\sb {xxx} \end{equation}
 after the substitution $\rho=-(1+q)\sp {-1/2}$.

Equation~(\ref{bhd1}) was discovered in an unpublished work by
Harry~Dym \cite{kruskalHD}, and appeared in a more general form in works
of P.~C.~Sabatier \cite{sabatierHD1,sabatierHD2,sabatierHD3}.
More recently, its relations with the
Kadomtsev-Petviashvili (KP) and modified-KP hierarchy have been studied in
detail by Oevel and Carillo~\cite{OevelCarillo}.

In the present work we discuss the HD hierarchy from the bi-Hamiltonian point
of view and show that it is amenable to the systematic treatment
developed  in \cite{CFMPZ98,FMP98,FLP,pondi,MPZ97}.
The main result of this paper is the existence of a {\em matrix HD hierarchy\/}, giving rise to the usual (scalar) HD hierarchy after a projection. It is well known that a lot of integrable PDEs can be obtained as suitable reductions from (integrable) hierarchies living on
loop-algebras, the main example being the Drinfeld-Sokolov reduction \cite{DS}. However, only recently it was realized that also the Camassa-Holm hierarchy has this important property \cite{FLP}. The HD case is settled in this work, starting from the results in \cite{LoPe}, where it has been shown that the bi-Hamiltonian structure of HD is the 
reduction of a suitable bi-Hamiltonian structure on the space $\CM=C^\infty(S^1,\mathfrak{sl}(2))$
of $C^{\infty}$ maps from the unit circle to $\mathfrak{sl}(2)$. 

The plan of this article is the following:

In Section~\ref{recall}
we review the general definitions of Poisson geometry and bi-Hamiltonian
theory. We review the important concept of an $\CS$-hierarchy which
was already used in \cite{Pbou} in connection with the Boussinesq equation.

Section~\ref{loop}
is devoted to endowing the loop-space on the Lie algebra of
traceless $2\times 2$ real matrices with a bi-Hamiltonian structure
following a construction in \cite{LoPe}.

Section~\ref{matrixHD}
describes the construction of the matrix HD hierarchy, i.e., a hierarchy
of commuting Hamiltonian flows in two fields that reduces to the
Harry Dym equation~(\ref{bhd1}) upon a suitable reduction.
Two-component extensions of the HD equation have interested a number of researchers,
see, e.g., \cite{af1988,aratyn2006,pop2003}.
We will see at the end of Section 5 that the hierarchy presented herein is different from 
those presented by these authors. 

We conclude 
with a Riccati-type equation for the conserved quantities of
the matrix HD hierarchy, in Section~\ref{riccati}, and a discussion concerning a reciprocal transformation 
in Section~\ref{recipr}.

\section{Bi-Hamiltonian preliminaries}
\label{recall} 

This section collects a number of facts from bi-Hamiltonian geometry. More information could be found in \cite{pondi}.

A \varb  \ is a triple $(\CM,P_1,P_2)$ consisting of a manifold $\CM$ and of two compatible Poisson tensors $P_1$ and $P_2$ on $\CM$.
In this context, we fix a symplectic leaf $\CS$ of $P_1$ and consider the distribution
$D=P_2(\mbox{Ker}P_1)$ on $\CM$. 
As it turns out, the distribution $D$ is integrable. Furthermore, if $E=D\cap T\CS$ is the distribution induced by $D$ on $\CS$ and the quotient space $\CN=\CS/E$ is a manifold, then it is a \varb.
In situations where an explicit description of the quotient manifold $\CN$ is not readily
 available, the following technique to compute the reduced \bih\
 structure  is very useful \cite{CP}.
 Assume that $\CQ$ is a submanifold of $\CS$ that is transversal to the distribution $E$, in the sense that
\begin{equation}
\label{split}
T_p\CQ\oplus E_p=T_p\CS\qquad\mbox{for all $p\in\CQ$}\ .
\end{equation}
Then, $\CQ$ is locally diffeomorphic to $\CN$ and inherits a \bih\ structure from $\CM$. 
The reduced Poisson pair on $\CQ$ is given by
\begin{equation}
\label{redP}
\left(P_i^{\mbox{\scriptsize rd}}\right)_{p}\al=\Pi_p\left((P_i)_p\tilde\al\right)\ ,\qquad i=1,2\ ,
\end{equation}
where $p\in\CQ$, $\al\in T^*_p\CQ$, the map $\Pi_p:T_p\CS\to T_p\CQ$ is the projection relative to \rref{split}, and $\tilde\al\in T^*_p\CM$ satisfies
\begin{equation}
\tilde\al|_{D_p}=0\ ,\qquad \tilde\al|_{T_p\CQ}=\al\ .
\end{equation}

Let us assume that $\{H_j\}_{j\geq 0}$ is a bi-Hamiltonian hierarchy on $\CM$, that is,
$P_2 dH_j=P_1 dH_{j+1}$ for all $j\geq 0$ and $P_1 dH_0=0$.  In other words, 
$H(\lambda)=\sum_{j\geq 0}H_j\lambda^{-j}$ is a (formal) Casimir of the {\em Poisson pencil\/} $P_2-\lambda P_1$. The bi-Hamiltonian vector fields associated with the hierarchy can be reduced on the quotient manifold $\CN$ according to
\begin{prop}
\label{reduhie}
The  functions $H_j$ restricted to $\CS$ are constant along the distribution
$E$. Thus,  they give rise to functions on $\CN$.
Such functions form a bi-Hamiltonian hierarchy with respect to the reduced Poisson pair. The vector fields $X_j=P_2dH_{j}=P_1dH_{j+1}$ are tangent to
$\CS$ and project on $\CN$. Their projections are the vector fields associated with the reduced hierarchy.
\end{prop}

In the sequel, we shall need a more general
definition than that of a  bi-Hamiltonian hierarchy. The point being that, once we
have fixed a symplectic leaf $\CS$ of $P_1$, it is not always
possible to determine a hierarchy on $\CM$ that is defined also on
$\CS$. In other words, there exist singular leaves for the 
hierarchies of a \varb. Nevertheless, it is sometimes possible to define
hierarchies which are, in a certain sense ``local'' on $\CS$. 
\begin{defi}
An {\em $\CS$--hierarchy\/} is a sequence $\{V_j\}_{j\geq 0}$ of maps from $\CS$ to $T^*\CM$,
$$
V_j:s\mapsto V_j(s)\in T_s^*\CM\ ,
$$
with the following properties:
\begin{itemize}
\item $V_j$ restricted to $T\CS$ is an exact $1$-form, that is,
there exist functions $H_j$ on $\CS$ such that $V_j|_{T\CS}=dH_j$;
\item $P_2V_j=P_1V_{j+1}$ for all $j\geq 0$ and $P_1 dH_0=0$.
\end{itemize}
 \end{defi}

Obviously, every bi-Hamiltonian hierarchy on $(\CM,P_1,P_2)$ defined in a neighborhood of $\CS$ gives
rise to an $\CS$--hierarchy. In contradistinction, in this paper we will see 
an example of $\CS$--hierarchy that does not come from any bi-Hamiltonian hierarchy. This is also the case of the Boussinesq hierarchy \cite{Pbou}.

It is not difficult to extend Proposition~\ref{reduhie} to the case of $\CS$--hierarchies. In the sequel, whenever talking about $\CS$--hierarchies and referring to such result, it shall be understood that we mean such straightforward extension. 

\section{A bi-Hamiltonian structure on a loop-algebra} \label{loop}

In this section we recall from \cite{LoPe} that the bi-Hamiltonian structure of the (usual) HD hierarchy can be obtained by means of a reduction.

Let $\CM = C^{\infty}(S^1,\mathfrak{sl}(2))$ be the loop-space on the Lie algebra of traceless $2\times 2$ real matrices, i.e., the space of $C^{\infty}$ functions from the unit circle $S^1$ to $\mathfrak{sl}(2)$.
 The tangent space $T_S\CM$ at $S\in \CM$  is identified with $\CM$ itself,
 and we will assume that $T_S\CM \simeq T^*_S\CM $ by the non-degenerate form
$$ \langle V_1, V_2\rangle = \int\mathrm{tr} (V_1(x)V_2(x))\, \di x,
 \qquad V_1,V_2 \in \CM\ , $$
where the integral is taken (here and throughout this article) on $S^1$. 
It is well-known \cite{LM} that the manifold $\CM$ has a $3$-parameter family of compatible Poisson tensors. To wit, 
\begin{equation}
\label{pencil}
 P_{(\lambda_1, \lambda_2, \lambda_3)}=\lambda_1 \partial_x
+\lambda_2[\,\cdot\, ,S] +\lambda_3 [\,\cdot\,
, A]\ ,
\end{equation}
where $\lambda_1, \lambda_2, \lambda_3 \in \RR$,
the matrix $A\,  \in \, \mathfrak{sl}(2)$ is  constant, and
\begin{equation*}
S=\begin{pmatrix} p & q\\ r & -p\end{pmatrix}\in\CM\ .
\end{equation*}
In this paper we focus on the pencil
\begin{equation}
\label{poipen} 
P_{\la}=P_2-\lambda P_1= \partial_x  +[\,\cdot \, , \, A+\lambda S]
\end{equation}
with
$$
A=\left( \begin{array}{cc}
0&0\\
1&0
\end{array}\right)\ . 
$$
This means that
\begin{equation}
\label{poistru} 
P_2=\partial_x  +[\,\cdot \, , \, A]\ ,\qquad P_1=[S, \,\cdot \,]\ .
\end{equation}
\begin{rem}
\label{remkdv}
We briefly recall from \cite{pondi} the construction of the matrix KdV hierarchy, since what we will do in the following for the HD case is completely similar, even though technically more complicated. In the KdV case the Poisson pair is 
\begin{equation*}
P_2=\partial_x  +[\,\cdot \, , \, S]\ ,\qquad P_1=[A, \,\cdot \,]\ 
\end{equation*}
and the symplectic leaf is chosen to be
\begin{equation*}
\CS=\left\{\begin{pmatrix} p & q \\ 1 & -p \end{pmatrix}
\mid p,q\in C^\infty(S^1,\mathbb{R})\right\}\ .
\end{equation*}
The quotient space $\CN$ can be identified with $C^\infty(S^1,\mathbb{R})$ and the projection from $\CS$ to $\CN$ is given by 
\begin{equation}
\label{projkdv}
(p,q)\mapsto u=p_x+p^2+q\ . 
\end{equation}
The matrix KdV hierarchy is given by the flows 
\begin{equation*}
\frac{\partial S}{\partial t_j}= P_2 V_{j-1}=P_1V_{j}
\end{equation*}
on the symplectic leaf $\CS$, where the $V_j$ are the coefficients of 
$$V(\la)=\sum_{j\ge -1} V_j \la^{-j}$$ and $V(\la)$ is uniquely determined by the conditions
$P_\la V(\la)=0$ and $\tr V(\la)^2=2\lambda$. The conserved densities can be found also by solving the Riccati equation $h_x+h^2=p_x+p^2+q+\lambda$. The usual (scalar) KdV hierarchy lives on the quotient space $\CN$ and can be obtained from the matrix one by applying the projection \rref{projkdv}. In this case, this amounts to the well known Drinfeld-Sokolov
reduction.

We close this remark by recalling that the Hamiltonian flows
$$ \partial_t S  =\partial_x V+ [V,S]\ , \quad V=\di H\ , $$ of the Poisson 
tensor $P_2=\partial_x+[\,\cdot \, , \, S]$
admit the zero-curvature representation
$$  [\partial_t+ V, \partial_x+ S]=0.$$
Such a representation does not seem to exist in our (HD) case.
\end{rem}

In \cite{LoPe} the bi-Hamiltonian reduction procedure was applied to
the pair $(P_1,P_2)$. In this case, 
\begin{equation*}
D_S=\left\{\begin{pmatrix}
(\mu p)_x+ \mu q & (\mu q)_x \\
(\mu r)_x-2 \mu p & -(\mu p)_x -  \mu q
\end{pmatrix}
\mid \mu\in C^\infty(S^1,\mathbb{R})\right\}\ ,\qquad S\in\CM .
\end{equation*}
The distribution $D$ is not tangent to the generic symplectic leaf of
 $P_1$. However, it is  tangent to the symplectic leaf
\begin{equation}
\label{symleaf}
\CS=\left\{\begin{pmatrix} p & q \\ r & -p \end{pmatrix}
\mid p^2+qr=0, (p,q,r)\not=(0,0,0)\right\}\ ,
\end{equation}
so that $E_p=D_p\cap T_p\CS$ coincides with $D_p$ for all $p\in\CS$.
It is not difficult to prove that the submanifold
\begin{equation}
\CQ=\left\{S(q)=\begin{pmatrix} 0 & q \\ 0 & 0 \end{pmatrix}
\mid q\in C^\infty(S^1,\mathbb{R}), q(x)\ne 0\ \forall x\in S^1\right\}
\end{equation}
of $\CS$ is transversal to the distribution $E$ and that the projection
$\Pi_{S(q)}:T_{S(q)}\CS\to T_{S(q)}\CQ$ is given by
\begin{equation}
\label{proj}
\Pi_{S(q)}:(\dot p,\dot q)\mapsto (0,\dot q-\dot p_x)\ .
\end{equation}
The reduced bi-Hamiltonian structure (\ref{redP}) coincides with the bi-Hamiltonian structure
of the Harry Dym hierarchy (see \cite{LoPe} for details):
\begin{eqnarray*}
&&\left(P_1^{\mbox{\scriptsize rd}}\right)_{q} =-(2q\partial_x+q_x)\\
&&\left(P_2^{\mbox{\scriptsize rd}}\right)_{q} = - \frac12 \partial_x^3\ .
\end{eqnarray*}
Starting from the Casimir $\int\sqrt{q}\, \di x$ of $P_1^{\mbox{\scriptsize rd}}$, one constructs a bi-Hamiltonian hierarchy, which is called the HD hierarchy. We refer to \cite{PSZ} and the references therein for more details, and for a discussion about a ``KP extension'' of the HD hierarchy (see also \cite{OevelCarillo}).

\begin{rem}
We take this opportunity for correcting a mistake in \cite{PSZ}. Equation
(3.5) in that paper should be replaced with
\begin{equation}
K^{(2j+1)}=\la\left(-\frac{1}{2}(\la^{j} w)_{+,x}+k(\la^{j} w)_+\right)\ .
\end{equation}
The remaining of the paper is correct, up to minor changes.
\end{rem}
We consider now the bi-Hamiltonian hierarchies of the Poisson pair (\ref{poistru}).
According to the  bi-Hamiltonian theory, the computation of the flows of these hierarchies is divided in two steps.
\begin{enumerate}
\item  First we have to look for a 1-form $V(\la)=\sum V_j \la^{-j} $ that belongs to the kernel of the pencil $P_{\la}$. By construction the coefficients $V_j$ satisfy the relations $$ P_2V_j=P_1V_{j+1}.$$ By definition the flows of the hierarchy are
\begin{equation}
\label{flows} 
 \frac{\partial S}{\partial t_j}= P_2 V_{j-1}=P_1V_{j}.
\end{equation}
In the case of $\CS$--hierarchies the 1-form $V(\la)$ is defined only on a symplectic leaf $\CS$.
\item Then,  we have to verify that the 1-form is exact. If this is the case, the coefficients of the potential $H(\la)=\sum H_j \la^{-j}$, the so-called Casimir of the pencil, are the Hamiltonians of the flows (\ref{flows}). In the case of $\CS$--hierarchies the Hamiltonians are the coefficients of the potential of the restriction of $V(\la)$ to $T\CS$.
\end{enumerate}
We apply now this procedure to the Poisson pencil (\ref{poipen}). In particular, we shall study a 1-form $V(\la)$ defined only on the symplectic leaf $ \CS$ defined by 
(\ref{symleaf}) and the corresponding $\CS$--hierarchy.

Let us suppose that 
$$
V= \begin{pmatrix} \alpha & \beta\\ \gamma & -\alpha\end{pmatrix}
$$ 
is a solution of $P_{\la}V =0$, that is,{
\begin{equation}
\label{Pla} 
V_x+[V\, , A + \la S] =0\ ,
\end{equation}
and let us write the previous equation in componentwise form 
\begin{equation}\label{system}
\begin{cases}
& \al_x+ (\la r +1) \, \be-\la q \,\ga =0 \\
& \be_x+ 2 \la q  \,  \al -2 \la p \,\be =0 \\
& \ga_x+ 2\la p\, \ga -2 (\la r +1)\, \al =0
\end{cases}
\end{equation}
Upon expressing $\al$ and $\ga$ in terms of $\be$,
\begin{equation}
\label{ag}
\begin{cases}
&\al=\dfrac{1}{2q}\big(\, -\dfrac{\be_x}{\la}+2 \be p\, \big)\\
&\ga=-\dfrac{\be_{xx}}{2\la^2 q^2}+ \dfrac{\be_{x}}{\la q^2} \big( \,  p+\dfrac{q_x}{2\la q} \, \big)+\be \big ( \, \dfrac{p_x}{\la q^2}- \dfrac{q_x p}{\la q^3}+\dfrac{r}{q}+\dfrac{1}{\la q} \,  \big)
\end{cases}
\end{equation}
we find that $\be$  satisfies the equation 
\begin{eqnarray*}
-\dfrac{\be_{xxx}}{2 q^2 \la^2}+\dfrac{3 q_{x}\,\be_{xx}}{2 q^{3}\lambda^{2}} +\left( \dfrac{2}{q \lambda}+ \dfrac{2 p_x}{ q^2 \lambda} +\frac{2 r}{q} -\frac{3 q_{x}^2}{2q^{4}\lambda^{2}}-
\frac{2 q_{x} p}{q^{3}\lambda} +\dfrac{q_{xx}}{2 q^3 \la^2} +\dfrac{2p^2}{q^2}   \right) \be_{x} +
\\
+\left( \dfrac{r_x}{q} -\dfrac{q_x}{q^2 \la}   +   \frac{p_{xx}}{q^2 \la}-\frac{3 q_{x}p_{x}}{q^3 \la} + \frac{3 q_{x}^{2}p}{q^4 \la}-
\frac{q_{xx}p}{q^3 \la}-\frac{q_{x}r}{q^2}+\frac{2p p_{x}}{q^2}-\frac{2 p^2 q_{x}}{q^3} \right)\be = 0
\end{eqnarray*}
This equation can be rewritten as 
\begin{equation}\label{18}
\frac{1}{\be}\, \frac{\di}{\di x} (\alpha^{2}+\be \ga)=0\ .
\end{equation}
Indeed, it is a well-known consequence of equation (\ref{Pla}) that 
 the spectrum of $V$ does not depend on $x$, so that 
$\dfrac{\di}{\di x} \tr V^2 = 0$. Let us set
\begin{equation} \label{tr}
  \tr \frac{V^2}{2} = \alpha^{2}+\be \ga= F(\lambda)\ ,
\end{equation}
where $F(\lambda)_x=0$. Then,  the equation for $\be$ becomes
\begin{equation} \label{beta}
2 q \be_{xx}\be -q\be_x^2-2 q_x \be \be_x +4 (q_x p-q p_x -q^2)\be^2 \la -4q ( p^2+qr) \la^2\be^2 +4q^3 F(\la)\la^2 = 0\ .
\end{equation}

We now consider the possibility of finding a solution  $\be (\la)$ of (\ref{beta})  as a formal series expansion in (negative) powers of $ \lambda$, that is, $\be=\sum_{i=-1}^ {\infty} \be_i \lambda^{-i}$. In order to find the coefficients $\be_i $ recursively, we must equate the coefficients of the same degree in $\la$ starting from the highest order one. 
We choose $F(\la) $ to be a power of $\la$.
Let us suppose that 
$q(x)\neq 0$ for all $x$. Then,  it turns out that we have to distinguish the two cases:
\begin{itemize}
\item  If \  $p^2 +q\,r\neq 0 $, then the degree of $F(\la)$ has to be even;
\item  If \ $p^2 +q\,r = 0 $, then the degree of $F(\la)$ has to be odd.
\end{itemize}
We are interested in the latter case, in order to perform the reduction process described in Section~\ref{recall}. 
If the degree of $F(\la)$ is odd then there exists a solution $\beta$ expanded in a formal Laurent series only on the symplectic leaf $\CS$. Such a solution cannot be extended outside $\CS$
 because if  $p^2 +q\,r\neq 0 $, then the degree of $F(\la)$ has to be even.
This means that in this article, we will study a $\CS$-hierarchy on the symplectic leaf (\ref{symleaf}) that cannot be obtained from a bi-Hamiltonian hierarchy on the whole $\CM = C^{\infty}(S^1,\mathfrak{sl}(2))$.

The bi-Hamiltonian hierarchy corresponding to the former case  will not be considered here, although the proof of the exactness given in the next section can be easily adapted to this case.

\section{The matrix HD hierarchy}
\label{matrixHD}
In this section we will show that it is possible to find a solution 
\begin{equation}
 V= \sum_{i=-1}^ {\infty} V_i \lambda^{-i}\ = \sum_{i=-1}^ {\infty} 
\begin{pmatrix} \al_i & \be_i\\ \ga_i & -\al_i\end{pmatrix} \lambda^{-i}
\end{equation}
of equation (\ref{Pla}) at the points of the symplectic leaf $\CS$, such that every
$V_i$ restricted to $T \CS $ is an exact $1$-form.
This yields an $\CS$-hierarchy, to be called the matrix HD hierarchy. 
We will see that it projects to the usual (scalar) HD hierarchy. In particular, its second vector field projects to the HD equation. 

First of all, we restrict to the symplectic leaf $\CS$ and we use the now classical dressing transformation method 
\cite{ZS,DS,CMP2} to show that the matrix $V(\la)$ whose entries are given by the solution of 
\rref{beta} and \rref{ag} defines an $\CS$-hierarchy if $F(\la)$ does not depend on the point of $\CS$. Indeed, equation (\ref{tr}) implies that there exists a nonsingular matrix $K$ such 
that
\begin{equation*}
V(\lambda)=K\Lambda K^{-1}\ ,
\end{equation*}
where 
$$
\Lambda = \left( \begin{array}{cc}
0&1\\ F(\lambda) & 0 \end{array}\right)\ .
$$
We note that any other traceless matrix $\Lambda$  depending only on $\la$ and such that $\tr \frac{\Lambda^2}{2}=F(\la)$  is suitable for our purpose. Our choice simplifies the following computations.
Let us introduce
\begin{equation}
\label{M}
M= K^{-1}( S+\frac{A}{\la})K- \frac{1}{\lambda} 
K^{-1}K_x\ .
\end{equation}
Thus, we have the following:  
\begin{prop}
\label{propexa}
If $F(\lambda)$ does not depend on the point $S\in\CS$, then 
$V(\la)$ restricted to $T \CS $ is an exact $1$-form. More precisely, if $H:\CS\to\mathbb R$ is given by 
\begin{equation}
\label{H}
H(\la)= \int \tr (M \Lambda)\, \di x\ ,
\end{equation}
then $V|_{T \CS }= \di H $. 
\end{prop}
\begin{proof}
If $V$ is a solution of (\ref{Pla}), then
$$  
\dfrac{1}{\la} K^{-1}V_x K+ \dfrac{1}{\la} K^{-1}[V, 
 A+\la S] K= 0\ . 
$$
This in turn, implies that
$$
\dfrac{1}{\la} \Lambda_x+ [\Lambda, M] =0\ .
$$
Since $\La  $ does not depend on $x$, we have that $\La$ commutes 
with $M$. Therefore, for every tangent vector
$\dot S$ to the \syml\ $\CS$, we have
\begin{eqnarray*}
\langle \di H, \dot{S} \rangle & = & \int \tr (\dot{M} 
\Lambda) \, \di x \,  = \,
\int \tr (K^{-1}\dot{S}K \Lambda) + \tr ( [M, 
K^{-1}\dot{K}] \Lambda) \, \di x \\
  & = & \int   \tr (\dot{S} K \Lambda K^{-1})\, \di x
= \int \tr (\dot{S} V)\, \di x =\langle  V,\dot{S} \rangle 
\ ,
  \end{eqnarray*}
since $\int \tr ( [M, K^{-1}\dot{K}] \Lambda)\, \di x=0$.
This completes the proof.
\end{proof}

\vspace{0.5truecm}
Let us now compute explicitly $H$. A possible choice for 
$K$ is
\begin{equation*}
K=\left( \begin{array}{cc}
  \be^{\frac{1}{2}} & 0 \\ -\al\be^{-\frac{1}{2}} & 
\be^{-\frac{1}{2}} \end{array}\right)\ \mbox{ .}
\end{equation*}
Since $M$ commutes with $\La$ and both matrices have distinct eigenvalues, it follows that $M$ is a 
polynomial of $\La$. However, since they are traceless and we are working with $2\times 2$ matrices it
follows that $M$ is a multiple of $\La$. This simplifies the computation of $M$, since it becomes
$$
M = \,\frac{q}{\be} \Lambda\ .
$$
Thus, we have that
\begin{equation}
\label{Hamilt}
H (\la) = \int 2 \frac{ q}{\be} F(\la) \,\di x\ .
\end{equation}
We define the {\em matrix HD hierarchy\/} to be the $\CS$-hierarchy corresponding to the choice $$F(\la)=\la.$$ In order to find its first vector fields, let us substitute $p^2+qr =0$ and $F(\la)= \la$ in equation (\ref{beta}), to find
\begin{equation} 
\label{beta2}
2 q \be_{xx}\be -q\be_x^2-2 q_x \be \be_x +4 (q_x p-q p_x -q^2)\be^2 \la  +4q^3\la^3 = 0\ .
\end{equation}
 From now on, we use the functions $p$ and $q$ to describe a point of $\CS$. 
We know that it is possible to solve equation \rref{beta2} recursively, starting from the highest power of $\la$:
\begin{eqnarray*}
 \la^3: & &  4(q_x p-q p_x-q^2) \be_{-1}^2 = -4 q^3\ .
\end{eqnarray*}
We choose the positive solution
\begin{equation}\label{be-1} 
  \be_{-1}=\sqrt{\dfrac{q^3}{q^2-q_x p + q p_x}}\ .
\end{equation}
 Using the expressions (\ref{ag}) for $\al$ and $\ga$ we get a recursive formula for the matrices $ V_{i}$. Indeed, we have that
\begin{equation}
\label{ag-1}
\begin{cases}
&\al_{-1}=p \dfrac{\be_{-1}}{q}\\
&\ga_{-1}=r  \dfrac{\be_{-1}}{q}
\end{cases}
\end{equation}
and 
\begin{equation}
\label{ag-ric}
\begin{cases}
&\al_i=\dfrac{1}{2q} (-(\be_{i-1})_x +2 \be_i p )\\
&\ga_i= \dfrac{1}{q} ( (\al_{i-1})_x+\be_i r + \be_{i-1}  )
\end{cases}
\end{equation}
for all $i\ge 0$. Therefore, we can compute the first $1$-form
$$
V_{-1} = \begin{pmatrix} p & q \\ r & -p \end{pmatrix} \varphi(x)\ ,
$$
where 
$$ \varphi(x) :=\sqrt{\dfrac{q}{q^2-q_x p+ q p_x}} \mbox{ .}$$
We can verify immediately that $V_{-1}$ indeed commutes with $S$, as expected.

Applying the Poisson tensor $P_2$ to $V_{-1}$ we obtain the first vector field 
$X_0:=P_2 (V_{-1}) = {V_{-1}}_x    + [V_{-1},A]$
of the hierarchy:
\begin{equation}
\begin{cases}
\label{chi0} 
&\dot{p}= (p \varphi)_x+ q \varphi\\
&\dot{q}= (q \varphi)_x 
\end{cases}
\end{equation}
We saw in Section~\ref{recall} that every $\CS$-hierarchy can be projected on the reduced 
bi-Hamiltonian manifold. 
We will show in Remark \ref{remprojHD} that the projection of the matrix HD hierarchy is the (scalar) HD hierarchy. Now we compute the projections of the first vector fields of the hierarchy. 
Since
$V_{-1}$ belongs to the kernel of $P_1$, we have that $V_{-1}|_{T\CS}=0$ and that $P_2(V_{-1})$ belongs to the distribution $D$, so that the projection of $X_0$ vanishes. 
However, let us check it explicitly. We must evaluate $X_0$ at the points $p=0$ of the transversal submanifold $\CQ$, then we have to project this vector field according to the formula  (\ref{proj}), thus obtaining the predicted result:  
\begin{equation*}
\frac{\partial q}{\partial t_0}=\dot{q}-\dot{p}_x=0\ .
\end{equation*}
More generally, 
let us observe that the formula for the vector field $X_i:=P_2 (V_{i-1})$ for $i\geq 0$  is 
\begin{equation}
\label{evolq}
\begin{cases}
&\dot{p}= {\alpha_{i-1}}_x+ \beta_{i-1}\\
&\dot{q}= {\beta_{i-1}}_x 
\end{cases}
\end{equation}
and its projection is $\dot{q}-\dot{p}_x=-{\alpha_{i-1}}_{xx}$ evaluated at $p=0$.

The next step in the iteration is:
\begin{eqnarray*}
\la^2: & &  2q \be_{-1}{\be_{-1}}_{xx}-q {(\be_{-1}}_{x})^2-2q_x \be_{-1}{\be_{-1}}_{x}  = 8(q^2-q_xp + qp_x) \be_{-1}\be_{0} \ .
\end{eqnarray*}
Using also equation (\ref{be-1}), we get that

\begin{eqnarray}\label{be0} 
\beta_0=
{\frac {\varphi  }{8\,q}\left( 2{q}^{2}\varphi\,\varphi_{xx}+2q{\varphi
}^{2}q_{xx}-{\varphi_{{x}}}^{2}{q}^{2}-3{\varphi}^{2}{q_{{x}}}^{2}
 \right)}
\end{eqnarray}
and then 
$$
V_{0} = \begin{pmatrix} \al_0 & \be_0 \\ \ga_0 & -\al_0  \end{pmatrix}\ , 
$$
where
\begin{eqnarray*}
\alpha_0&=&
-\frac{\varphi_x}{2}-\frac{q_x \varphi}{2q}+\frac{p\varphi}{8 \,q^2}
 \left( 2{q}^{2}\varphi\varphi_{{xx}}+2q{\varphi
}^{2}q_{xx}
-{\varphi_{x}}^{2}{q}^{2}-3{\varphi}^{2}{q_{x}}^{2}
 \right)
 \\
\gamma_0&=&\frac {1}{8\,{q}^{3}} \left(8\varphi_{{x}}p{q}^{2}+8\varphi p_{{x}}{q}^{2}-2{\varphi}^{
2}{p}^{2}{q}^{2}\varphi_{xx}-2{\varphi}^{3}{p}^{2}qq_{{xx}}+
\varphi{p}^{2}{\varphi_{{x}}}^{2}{q}^{2}+ \right.\\
&& \left. +\, 3{\varphi}^{3}{p}^{2}{q_{x}}^{2}
+8\varphi{q}^{3}\right) .
\end{eqnarray*}
We can now determine the second vector field $X_1:=P_2 (V_{0}) = {V_{0}}_x    + [V_{0},A]$. It is given by
\begin{equation}
\begin{cases}
\label{chi1}
&\dot{p}= -\dfrac{1}{2}\varphi_{xx}-\left(\dfrac{q_x}{2q}\varphi\right) _{x} + \left( \dfrac{p}{q} \be_0 \right)_{x} +\be_0 \\
&\dot{q}= {\be_0}_x 
\end{cases}
\end{equation}
Using equation (\ref{be0}), we can write out the above equation as follows:
\begin{equation}
\label{chi12}    
\left\lbrace 
\begin{array}{lll}
&\dot{p}=  &\! \left(
 \dfrac{\varphi}{8q} +\dfrac{p_x \varphi}{8q^2}-\dfrac{q_x\varphi}{8q^3} \right)
\left( 2{q}^{2}\varphi\,\varphi_{xx}+2q{\varphi
}^{2}q_{xx}-{\varphi_{x}}^{2}{q}^{2}-3{\varphi}^{2}{q_{{x}}}^{2}
\right)+\\
& &  
 -\dfrac{1}{2}\varphi_{xx}-\left(\dfrac{q_x}{2q}\varphi \! \right) _{x} \! \! + \dfrac{p}{q}\left( {\dfrac {\varphi  }{8q}\left( 2{q}^{2}\varphi \varphi_{xx} \! +2q{\varphi
}^{2}q_{xx}-{\varphi_{{x}}}^{2}{q}^{2}-3{\varphi}^{2}{q_{{x}}}^{2}
 \right)} \! \right) _{x} 
\\ \\
&\dot{q}=& \left( {\dfrac {\varphi  }{8\,q}\left( 2{q}^{2}\varphi\,\varphi_{xx}+2q{\varphi
}^{2}q_{xx}-{\varphi_{{x}}}^{2}{q}^{2}-3{\varphi}^{2}{q_{{x}}}^{2}
 \right)} \! \right) _{x}
\end{array}      \right.
\end{equation}
Starting from \rref{chi1}, we calculate the reduced vector field, first evaluating $X_1$ at the points $p=0$ of the transversal submanifold $\CQ$,
\begin{equation}
\begin{cases}
\label{chi1rid} 
&
\dot{p}= \dfrac {5\, {q_{x}}^{2}}{32\,{q}^{ \frac{5}{2}}}-\dfrac {q_{xx}}{8\, {q}^{\frac{3}{2}}}
\\  \\
&\dot{q}=  \left( - \dfrac {7\, {q_{x}}^{2}}{32\,{q}^{\frac{5}{2}}}+\dfrac {q_{
xx}}{8\, {q}^{\frac{3}{2}}} \right)_{x} 
\end{cases}
\end{equation}
and then projecting this vector field on the transversal submanifold. We thus obtain the 
HD equation~(\ref{bhd1}) 
\begin{equation*}
\frac{\partial q}{\partial t_1}=\dot{q}-\dot{p}_x=-\dfrac{1 }{2}\left( \dfrac{1}{\sqrt{q}}\right)_{xxx}\ \mbox{ .}
\end{equation*}
 This equation is equivalent to (\ref{bhd1}) after the change of 
 variables $q \mapsto (1 + q)$ and $t_1 \mapsto - 4 t$.

\section{A Riccati equation for the conserved densities} \label{riccati}

The goal of this section is to point out that the conserved densities of the matrix HD hierarchy can also be found by means of a Riccati-type equation. 

We recall that equation (\ref{Hamilt}) gives, for $F(\la)=\la$, the expression of the potential $H$ of the $1$-form $V|_{T\CS}$. The corresponding density is clearly defined up to a total $x$-derivative. This fact allows us 
to introduce
$$ 
h=\dfrac{q \la^\frac32}{\be}   + \dfrac{\be_x}{2 \be}\ \mbox{, }
$$ 
which transforms the equation (\ref{beta}) in the Riccati-type equation
\begin{equation} 
\label{ricc} 
 h_x+h^2-\dfrac{q_x}{q}h= \left( p_x+q- \dfrac{q_x}{q}p\right) \la\ .
\end{equation}
Its solution $h$ yields  
\begin{equation} 
\label{Hconh}
H(\la)=\frac{2}{\sqrt{\la}}\int h \di x\ 
\end{equation}
for the functional $H$.
We set $z=\sqrt{\la}$, and substitute $h=\sum_{i=-1}^{\infty} h_i z^{-i}$ in the Riccati 
equation (\ref{ricc}), which takes the form 
\begin{equation}
\sum_{i=-1}^{\infty} \Big( { h_{i}}_{x} +\sum_{j=0}^{1}(h_{i-j}h_{i})\Big) z^{-i} -\dfrac{q_x}{q}\sum_{i=-1}^{\infty} h_i z^{-i}= \left( p_x+q- \dfrac{q_x}{q}p\right) z^2 \ .
\end{equation}
Once again, this equation can be solved recursively, starting from the highest degree of $z$. The first step is
\begin{equation*}
 z^2: \qquad h_{-1}^{2}= p_x+q- \dfrac{q_x}{q}p\ \mbox{ ,}
\end{equation*}
which gives, up to a sign, 
\begin{equation*}
h_{-1}=\sqrt{p_x+q- \dfrac{q_x}{q}p}\ .
\end{equation*}
Similarly, we have that 
\begin{eqnarray*}
 z^1: & & {h_{-1}}_x+2h_{-1}h_{0}-\dfrac{q_x}{q} h_{-1}=0 \,
\end{eqnarray*}
from which we obtain
\begin{equation*}
h_{0}=-\dfrac{{h_{-1}}_x}{2 h_{-1}}+ \dfrac{q_x}{2q} \ .
\end{equation*}
Let us notice that this is a total $x-$derivative. 
More generally, it is evident from \rref{Hconh} that every even density is a total $x-$derivative. 
Indeed, $H(\la)=\sum_{i\ge 0}H_i\la^{-i}$, with  
\begin{equation} 
\label{Hconhi}
H_i=2\int h_{2i-1} \di x\ .
\end{equation}
In particular, $H_0=2\int h_{-1} \di x $, and it can be checked that $\di H_{0}= V_0|_{T\CS}$, as claimed in
 Proposition \ref{propexa}.

The next equation is
\begin{eqnarray*}
 z^0: & & {h_0}_{x} +2h_{-1} h_{1}+h_0^2-\dfrac{q_x}{q}h_0 =0
\end{eqnarray*}
and the corresponding density is
\begin{equation*}
h_{1}=-\dfrac{1}{2 h_{-1}} \left({h_0}_{x} + h_0^2+ \dfrac{q_x}{q} h_0\right)\ . 
\end{equation*}
This leads to 
\begin{equation*}
H_1= 2 \int \left(
\frac {{h_{-1}}_{xx}}{ 4 h_{-1}^{2}}-\frac{3{h_{-1}}_x^2}{8h_{-1}^3}  -\frac{q_{xx}}{4 h_{-1}q}+\frac{3q_x^2}{8h_{-1}q^2}
 \right)  \di x \ .
\end{equation*}
Integrating by parts and substituting the expression for $h_1$, we find
\begin{equation*}
H_1= 2 \int \left(
\dfrac {\left( p_{xx}+q_x-\big(\dfrac{p q_x}{q}\big)_x 
  \right)^2 }{32 \, \left( p_x+q-\dfrac{pq_x}{q} 
\right)^{\frac{5}{2}} } -\dfrac{q_{xx}}{4 q\, 
\sqrt{p_x+q-\dfrac{p q_x}{q}}} +\dfrac{3q_x^2}{8\,q^2 \, 
\sqrt{p_x+q-\dfrac{p q_x}{q}}}
  \right)  \di x\ .
\end{equation*}
This is the Hamiltonian (with respect to the symplectic structure obtained by restricting $P_1$ to its symplectic leaf $\CS$) of the $2$-component extension (\ref{chi12}) of the HD equation.

\begin{rem}
\label{remprojHD}
In this paper we chose to express the equation (\ref{tr}) in terms of $\be$, using the equations of the system (\ref{system}). An analogous calculation can be performed in terms of $\ga$, leading to the expression  
\begin{equation*}
 \tilde H(\la) = \int 2 \frac{\la r +1}{ \la \ga } F(\la)  \di x
\end{equation*} 
for the functional on $\CS$ such that $V|_{T \CS }= \di\tilde H $ (see equation (\ref{Hamilt}) and Proposition \ref{propexa}). The functional $H$ defined by (\ref{Hamilt}) could in principle differ from $\tilde H$ by an additive constant, but we will see that they coincide. In section \ref{matrixHD} we set $F(\la)=\la$ and then in this section we proved that 
$$
H(\la)=\frac{2}{\sqrt{\la}}\int h \di x\ ,
$$
where $h$ satisfies the Riccati-type equation (\ref{ricc}). In the same way
we can prove that
\begin{equation*}
\tilde H(\la)=\frac{2}{\sqrt{\la}}\int \tilde h \di x\ ,
\end{equation*}
where  $\tilde h$ satisfies the following different Riccati-type equation,
\begin{equation}
\label{ricc2}
({\tilde h}_x+{\tilde h}^2)(\la r+1)+\la r_x \tilde h = \la(q-p_x)+\la^2 ( r_x  p-p_x r+ p^2+2qr )\ ,
\end{equation}
where $p^2+q r= 0$ since we are on the symplectic leaf $\CS$.
The two Riccati equations are both equivalent to equation (\ref{tr}), respectively expressed in terms of $\be$ and $\ga $ through the transformations
$$ 
h=\dfrac{q \la^\frac32}{\be}   + \dfrac{\be_x}{2 \be}\ \mbox{, }
$$ 
and 
$$ 
\tilde h=\frac{\la r +1}{\ga}\la^\frac12+\frac{\ga_x}{2 \ga}.
$$
It can be checked directly that the densities $h$ and $\tilde h$ differ by a total $x$-derivative, so that $H$ and $\tilde H$ actually coincide. Indeed, using (\ref{tr}) and the system (\ref{system}), we obtain
\begin{eqnarray*}
h-\tilde h &=&  \frac{\sqrt{\la} \,\al_x}{\beta \ga} +\frac{\be_x}{2 \be}-\frac{\ga_x}{2 \ga} \\
&= &       \frac{\sqrt{\la}\, \al_x}{\la-\al^2} +\frac{\be_x}{2 \be}-\frac{\ga_x}{2 \ga} \\
& =&     \frac{1}{2}\, \partial_x \, \left( \sqrt{\la}\, \log \left|\frac{\sqrt{\la} +\al}{\sqrt{\la} -\al } \right|  + \log \left| \frac{\be}{\ga}\right| \right).
\end{eqnarray*}
The choice of dealing with the first Riccati equation (\ref{ricc}) is due to the fact that equation (\ref{ricc2}) is more complicated to handle, in order to perform the iteration. 
But it is easy to see that (\ref{ricc2}), evaluated at the points of  $\CQ$, is the Riccati equation for the scalar HD hierarchy (see, e.g., \cite{PSZ}),
$$ 
\tilde h_x+\tilde h^2= q \la\ .
$$
This shows that the matrix HD hierarchy projects on the usual HD hierarchy.
\end{rem}

We close this section with an explanation of the difference between our 2-component extension of the HD hierarchy and those already present in the literature. Ours is more precisely a {\em lifting\/}, since it gives rise to the usual HD hierarchy after a {\em projection\/}. On the contrary, those already known in the literature {\em restrict\/} to the HD hierarchy. For example, in \cite{af1988} one has to put one of the two fields equal to zero. This is
completely obvious if one looks at the corresponding second order linear
problems (from our point of view, the Riccati equations): 
our equation (\ref{ricc}) is similar to
the one of the usual HD hierarchy, while the one in \cite[eq.(20)]{af1988} is different, since a polynomial of degree 2 in $\lambda$ appears. The same happens for the matrix KdV hierarchy (see Remark \ref{remkdv}) and the $N$-component extensions discussed in \cite{af1987}. 

\section{A reciprocal transformation}
\label{recipr}
In this final section we consider a change of variables that remarkably simplifies the form of the equations
of the two-component hierarchy studied in this paper. This change of variables is composed by two transformations:
\begin{enumerate}
 \item 
A reciprocal transformation;
\item 
A Miura-type transformation.
\end{enumerate}

Let us briefly recall the definition of reciprocal transformations. For further information see, for example, \cite{FePa} and references therein. 
Let 
\begin{equation}
\label{ePDEs}
u^i_t=P^i(u,u_x,u_{xx},...),\,\,\,\qquad i=1,\dots,n,
\end{equation}
be a system of evolutionary PDEs, where $u=(u^1,\dots,u^n)$.  A reciprocal transformation is a change 
of the independent variables $(x,t)\mapsto(\tilde{x},\tilde{t})$ of the form
\begin{eqnarray*}
d\tilde{x}&=&h(u)\,dx+k(u)\,dt\\
d\tilde{t}&=&f(u)\,dx+g(u)\,dt.
\end{eqnarray*}
For the purposes of the present work it suffices to consider the subclass of reciprocal transformations of the form
\begin{equation}
\label{RT}
d\tilde{x}=h(u)\,dx+k(u)\,dt,\,\,\,\,\,\,d\tilde{t}=dt.
\end{equation}
Clearly, the transformation (\ref{RT}) is locally well defined if, and only if, the right hand-side(s) are closed 
$1$-forms. This implies that $h$ and $k$ are the density and the  current of a conservation law
 of the system (\ref{ePDEs}). 

Suppose now that the system (\ref{ePDEs}) belongs to a hierarchy of commuting flows. In this case 
 the reciprocal transformation (\ref{RT}) has the form
\begin{eqnarray}
\label{times} d\tilde{x}&=&h\,dx+k_0\,dt_0+k_1\,dt_1+k_2\,dt_2+k_3\,dt_3+k_4\,dt_4+\dots\\
d\tilde{t}_i&=&dt_i,\,\,\,\qquad i=0,1,2,3,4,\dots
\end{eqnarray}
where $t_0,t_1,t_2,t_3,t_4,...$ are the times of the hierarchy, $h$ is a density of a conservation law
 and $k_0,k_1,k_2,k_3,k_4,...$ are the currents associated to the different times of the hierarchy.

Let us now focus our attention on the two-component hierarchy studied in this paper. 
In this case, the second equation of \rref{evolq} implies that $q$ is a density of a conservation law and $\beta_{i-1}$ is the current associated to the time $t_i$. Therefore, we can define the reciprocal transformation
\begin{eqnarray*}
dz &=&q\,dx+\beta_{-1}\,dt_0+\beta_0\,dt_1+\beta_1\,dt_2+\dots\\
d\tilde{t}_i&=&dt_i,\,\,\,i=0,1,2,3,4,\dots.
\end{eqnarray*}
Combining this transformation with the Miura-type transformation $(p,q)\mapsto(u,v)$
\begin{equation*}
u=\frac{1}{q} +\left(\frac{p}{q}\right)_z\ ,\qquad
v=\frac{1}{q}\ ,
\end{equation*}
the flows of the hierarchy assume a triangular form. Let us explain this fact in details. First of all, we observe that the Riccati equation (\ref{ricc}) in the new independent variables $(z,t)$
 and in new dependent variables $(u,v)$ becomes the usual Riccati equation for the Harry Dym hierarchy,
\begin{equation}
\chi_z + \chi^2 = \lambda u, 
\end{equation}
where $\chi = hv$. This remark suggests that the variable $u$ should evolve according to the Harry Dym hierarchy.
This fact can be verified as follows: We start by observing that
\begin{equation}
\label{udyn}
u_{t_i}=-\frac{q_{t_i}}{q^2}\left(1+p_z-2\frac{pq_z}{q}\right)
-\frac{q_z p_{t_i}}{q^2}+\frac{(p_{t_i})_z}{q}-\frac{p(q_{t_i})_z}{q^2}.
\end{equation}
In order to write the equation for the variable $u$ we need to know the evolution of
 $p$ and $q$ in the new independent variables $(z,t)$. 
Taking into account the additional contributions $-\beta_{i-1}q_z$ and  $-\beta_{i-1}p_z$, due to
 the dependence of the new independent variable $z$ on $t$, it turns out that $q$ and $p$ evolve 
according to the equations
\begin{eqnarray*}
q_{t_i} &=& q (\beta_{i-1})_z -\beta_{i-1}q_z\\
p_{t_i}&=& q\left(-\frac1{2}(\beta_{i-2})_{z}+\frac{p}{q}\beta_{i-1}\right)_z
+\beta_{i-1}-p_z\beta_{i-1}\\
&=&-\frac12 q (\beta_{i-2})_{zz}+p(\beta_{i-1})_z+\left(1-\frac{pq_z}{q}\right)\beta_{i-1} \mbox{, }
\end{eqnarray*}
where the first equation of \rref{ag-ric} has also been used.
Substituting these expressions in (\ref{udyn}), we obtain the equations for the variable $u$, namely
$$
u_{t_i}=-\frac{1}{2}\left(\beta_{i-2}\right)_{zzz}.
$$
The functions $\beta_{i}$ are polynomials in the derivatives of $u$ and can be recursively reconstructed from the equation
\begin{equation*}
\chi = \frac{\lambda^{3/2}}{\beta} + \frac{\beta_z}{2 \beta}\ .
\end{equation*}
For instance,
\begin{equation*}
\beta_{-1}=\frac{1}{\sqrt{u}},\,\quad\beta_0=\frac{1}{4}\beta_{-1}^2(\beta_{-1})_{zz}-
\frac{1}{8}\beta_{-1}[(\beta_{-1})_z]^2,
\,\dots.
\end{equation*}
Since $\beta_{i}=0$ for $i\leq -2$, we have that
\begin{eqnarray*}
u_{t_0}&=&0\\
u_{t_1}&=&-\frac{1}{2}\left(\frac{1}{\sqrt{u}}\right)_{zzz}.
\end{eqnarray*}
Summarizing, the two-component hierarchy studied in this paper is related by a reciprocal transformation to the triangular hierarchy
\begin{eqnarray*}
v_{t_i} &=& -(v\beta_{i-1})_z \\
u_{t_i} &=& X_i(u,u_z,u_{zz},\dots)
\end{eqnarray*}
where $X_i$ is the $i$-th flow of the Harry Dym hierarchy and $\beta_{i}$ is a polynomial in $u$ and its derivatives. The geometrical counterpart of this remark is clearly the fact (showed in the previous section) that the matrix HD hierarchy projects onto the usual HD hierarchy.

\section*{Acknowledgments}
The authors are grateful to Giovanni Ortenzi for fruitful discussions, and to an anonymous referee for suggesting us the change of variables considered in Section~\ref{recipr}. M.~P. and J.~P.~Z. would like to thank for the hospitality the Department  {\em Matematica e Applicazioni\/} of the Milano-Bicocca  University, where most of this work was done.

This work has been partially supported by the European Community through the FP6 Marie Curie RTN {\em ENIGMA} (Contract number MRTN-CT-2004-5652) and by the European Science Foundation through the {\em MISGAM}
program. J.P.Z was supported by CNPq grants 302161/2003-1 and
474085/2003-1.

\end{document}